\documentclass[12pt, twocolumn,showpacs,preprintnumbers,aps,prl,reprint,superscriptaddress]{revtex4-1}

\usepackage[pdftex]{graphicx}
\usepackage{subfigure}
\usepackage{color}
\usepackage{amsmath}
\usepackage[linktocpage,colorlinks=true,linkcolor=blue,citecolor=blue,breaklinks=true]{hyperref}
\usepackage{verbatim}
\usepackage{breakcites}

\begin{document}

\preprint{}

\title{Biphasic, Lyotropic, Active Nematics}

\author{Matthew L. Blow}
\affiliation{Centro de F\'{i}sica Te\'{o}rica e Computacional, Avenida Professor Gama Pinto 2, P-1649-003 Lisboa, Portugal.} 
\affiliation{Departamento de F\'{i}sica da Faculdade de Ci\^{e}ncias, Universidade de Lisboa, P-1749-016 Lisboa, Portugal.}

\author{Sumesh P. Thampi}
\affiliation{The Rudolf Peierls Centre for Theoretical Physics, 1 Keble Road, Oxford, OX1 3NP, UK}

\author{Julia M. Yeomans}
\affiliation{The Rudolf Peierls Centre for Theoretical Physics, 1 Keble Road, Oxford, OX1 3NP, UK}
\email[j.yeomans1@physics.ox.ac.uk]{j.yeomans1@physics.ox.ac.uk}
\homepage[]{https://www-thphys.physics.ox.ac.uk/people/JuliaYeomans/}

\date{\today}

\begin{abstract}

We perform dynamical simulations of a two-dimensional active nematic fluid in coexistence with an isotropic fluid. Drops of active nematic become elongated, and an effective anchoring develops at the nematic-isotropic interface.  The activity also causes an undulatory instability of the interface. This results in defects of positive topological charge being ejected into the nematic, leaving the interface with a diffuse negative charge. Quenching the active lyotropic fluid results in a steady state in which phase-separating domains are elongated and then torn apart by active stirring.

\end{abstract}


\maketitle

Many biophysical nematic systems, including microtubule bundles~\cite{Dogic2012}, cytoskeletal filaments ordered by molecular motors in motility assays~\cite{KierfeldFrentzel}, actin filaments~\cite{ChakrabartiDas}, cells~\cite{Poujade2007, Petitjean2010} and dense suspensions of microswimmers~\cite{Julia2012} are active, meaning that the constituent particles generate motion by dissipating chemical energy, for example from adenosine tirphosphate~\cite{Sriram2010, Marchetti2013}. This motion collectively manifests itself as a stress that keeps the system out of equilibrium. Active nematics exhibit rich pattern formation~\cite{GiomiMahadevan,Giomi2013,ThampiYeomans} and collective motion~\cite{TjhungMarenduzzo}.

Almost all studies of active nematics thus far have concentrated on bulk systems. However there are many examples where active material coexists with an isotropic fluid. These include active droplets~\cite{Dogic2012, TjhungMarenduzzo, GiomiDeSimone, Joanny2012}, biofilms, bacterial colonies and bacterial carpets~\cite{Wilking2011, Boyer2011, Darnton2004}. Existing studies include spontaneous division and motility of active nematic droplets through self generated flows~\cite{GiomiDeSimone}. However, we are not aware of research reporting the phase separation of active fluids nor the role of topological defects in lyotropic active nematics.

Therefore in this Letter we describe the behaviour of active nematic - isotropic mixtures. We show that active forces lead to nematic anchoring at the interface, as observed in growing bacterial colonies~\cite{Halletschek2007,Volfson2008}. Moreover, active forces elongate nematic domains in the direction parallel (perpendicular) to the director field for extensile (contractile) systems. The elongated domains are torn apart by hydrodynamic instabilites, which balance the tendency to phase ordering, forming a dynamic steady state with characteristic length scales. Furthermore, we find that defect formation is dominated by the ejection of point defects with topological charge $+1/2$ from the interface, leaving the interface itself with a negative topological charge.

The nematic order of the fluid is described by a symmetric, traceless tensor $\mathbf{Q}$~\cite{DeGennesBook}. We assume that the director always remains within the plane of the system so $Q_{\alpha\beta}=S\left(2n_{\alpha}n_{\beta}-\delta_{\alpha\beta}\right)$ is two-dimensional, with $\mathbf{n}$ the director and $S$ the magnitude of the order. The active nematic fluid is mixed with an isotropic fluid, and the amount of each is conserved. We use a scalar parameter $\phi$ to measure the relative density of each at a given point. The free energy of the system is 
\begin{equation}
\mathcal{F}=\int\left(f(\mathbf{Q},\mathbf{\nabla Q},\phi,
\mathbf{\nabla}\phi)-\mu\phi\right)d^{2}\mathbf{r},
\label{eqn:freeenergy}
\end{equation}
with
\begin{eqnarray}
f&=&\tfrac{1}{2}A\phi^{2}\left(1-\phi\right)^{2} + \tfrac{1}{2}C\left(S_{\mathrm{nem}}^{2}\phi-\tfrac{1}{2}Q_{\alpha\beta}Q_{\alpha\beta}\right)^{2} \nonumber\\
&+&\tfrac{1}{2}K\partial_{\gamma}\phi\partial_{\gamma}\phi + \tfrac{1}{2}L\partial_{\gamma}Q_{\alpha\beta}\partial_{\gamma}Q_{\alpha\beta},     \label{eqn:freeEnergy}\\
\mu&=&\frac{\partial f}{\partial\phi}-\partial_{\gamma}\left(\frac{\partial f}{\partial (\partial_{\gamma}\phi)}\right)  
\end{eqnarray}
where $A$, $C$, $K$ and $L$ are positive constants. The first term in $f$ is the bulk energy of the binary fluid~\cite{ChaikinBook,Orlandini1995}, which has two equilibria at $\phi=0,1$. The second term is the bulk energy of the liquid crystal~\cite{DeGennesBook}, and here it couples $S$ to $\phi$. We note that $Q_{\alpha\beta}Q_{\alpha\beta}=2S^{2}$, and thus isotropic order is favoured in regions where $\phi=0$, while nematic ordering with $S=S_{\mathrm{nem}}$ is favoured where $\phi=1$. This biphasic bulk energy permits a diffuse interface when combined with the third and fourth terms, which penalise gradients in $\phi$ and $\mathbf{Q}$ respectively~\cite{Orlandini1995,Sulaiman2006}. Both terms contribute to the surface tension of this interface, and the fourth term also provides the nematic elasticity in the bulk. $\mu$ is a Lagrange multiplier that conserves the integrated value of $\phi$.

The order parameters $\phi$ and $\mathbf{Q}$ evolve according to the convection-diffusion equations~\cite{Cahn1958,Berisbook}
\begin{align}
\partial_{t}\phi+\partial_{\beta}(\phi u_{\beta})&=M\nabla^{2}\mu, \label{eqn:cahnHilliard} \\\left(\partial_{t}+u_{\kappa}\partial_{\kappa}\right)Q_{\alpha\beta}&=-\zeta\Sigma_{\alpha\beta\kappa\lambda}\Lambda_{\kappa\lambda}-\mathrm{T}_{\alpha\beta\kappa\lambda}\Omega_{\kappa\lambda}+\Gamma H_{\alpha\beta} \label{eqn:berisEdwards}.
\end{align}
On the rhs of Eq.~(\ref{eqn:berisEdwards}) the first two terms form the upper convected derivative, which accounts for the rotation of the nematic under shear. $\zeta$ is the tumbling parameter and 
\begin{eqnarray}
\Lambda_{\alpha\beta}&=&\tfrac{1}{2}\left(\partial_{\beta}u_{\alpha}+\partial_{\alpha}u_{\beta}\right),
\Omega_{\alpha\beta}=\tfrac{1}{2}\left(\partial_{\beta}u_{\alpha}-\partial_{\alpha}u_{\beta}\right),\nonumber\\
\Sigma_{\alpha\beta\kappa\lambda}&=&S_{\mathrm{nem}}^{-1}Q_{\alpha\beta}Q_{\kappa\lambda}-\delta_{\alpha\kappa}\left(Q_{\lambda\beta}+S_{\mathrm{nem}}\delta_{\lambda\beta}\right)\nonumber\\
&-&\left(Q_{\alpha\lambda}+S_{\mathrm{nem}}\delta_{\alpha\lambda}\right)\delta_{\kappa\beta}+\delta_{\alpha\beta}\left(Q_{\kappa\lambda}+S_{\mathrm{nem}}\delta_{\kappa\lambda}\right),\nonumber\\
\mathrm{T}_{\alpha\beta\kappa\lambda}&=&Q_{\alpha\kappa}\delta_{\beta\lambda}-\delta_{\alpha\kappa}Q_{\beta\lambda}.
\end{eqnarray}
The final term describes the relaxation of $\mathbf{Q}$ to the minimum of the free energy:
\begin{multline}
H_{\alpha\beta}=\tfrac{1}{2}\left(\delta_{\alpha\beta}\delta_{\kappa\lambda}-\delta_{\alpha\kappa}\delta_{\beta\lambda}-\delta_{\alpha\lambda}\delta_{\beta\kappa}\right)\\
\times\left\{\frac{\partial f}{\partial Q_{\kappa\lambda}}-\partial_{\gamma}\left(\frac{\partial f}{\partial (\partial_{\gamma}Q_{\kappa\lambda})}\right)\right\}.
\end{multline}

The total density $\rho$ and the velocity $\mathbf{u}$ obey 
\begin{eqnarray}
&&\hspace{-0.5cm}\partial_{t}\rho+\partial_{\beta}(\rho u_{\beta})=0,      \label{eqn:continuity} \\
&&\hspace{-0.5cm}\partial_{t}(\rho u_{\alpha})+\partial_{\beta}(\rho u_{\alpha}u_{\beta})=-b u_{\alpha}+ \partial_{\beta}
\Big(2\rho\eta\Lambda_{\alpha\beta}-p_{0}\delta_{\alpha\beta}\nonumber\\
&&\;\;\;\;\;\;\;\;\;\;\;\;+\Pi_{\alpha\beta}+\left\{\zeta\Sigma_{\alpha\beta\kappa\lambda}+\mathrm{T}_{\alpha\beta\kappa\lambda}\right\}H_{\kappa\lambda}-\chi Q_{\alpha\beta}\Big), \label{eqn:navierStokes} 
\end{eqnarray}
where $b$ is a drag coefficient accounting for friction from the substrate. There are four passive contributions to the stress on the rhs of Eq.~(\ref{eqn:navierStokes}). The first and second are the usual Newtonian stress, with $\eta$ the kinematic viscosity and $p_{0}=\rho/3$ the isotropic pressure. The third and fourth are elastic stresses, with
\begin{equation}
\Pi_{\alpha\beta}=(f-\mu\phi)\delta_{\alpha\beta} - \frac{\partial f}{\partial(\partial_{\beta}\phi)}\partial_{\alpha}\phi - \frac{\partial f}{\partial(\partial_{\beta}Q_{\kappa\lambda})}\partial_{\alpha}Q_{\kappa\lambda}.
\end{equation}
The final term is the active stress. $\chi$ is the strength of activity, and a positive (negative) $\chi$ corresponds to an extensile (contractile) material~\cite{Sriram2002,Joanny2005,Davide2007}.

We solve the equations of motion using a hybrid lattice Boltzmann method. This involves solving Eqs.~(\ref{eqn:cahnHilliard}-\ref{eqn:berisEdwards}) using finite difference methods, and Eqs.~(\ref{eqn:continuity}-\ref{eqn:navierStokes}) using a lattice Boltzmann algorithm~\cite{Davide2007,TjhungMarenduzzo}. Parameters used are $S_{\mathrm{nem}}=1$, $A=0.08$, $C=0.5$, $L=0.005$, $K=0.01$, $\Gamma=0.1$, $M=0.1$, $\eta=1/6$, $b=0.1$, $\zeta=0.3$ (tumbling regime) and on average $\rho=40$. For these parameters, the characteristic interface width is $\sim 3$ lattice spacings. 

\begin{figure}
\includegraphics[width=\linewidth]{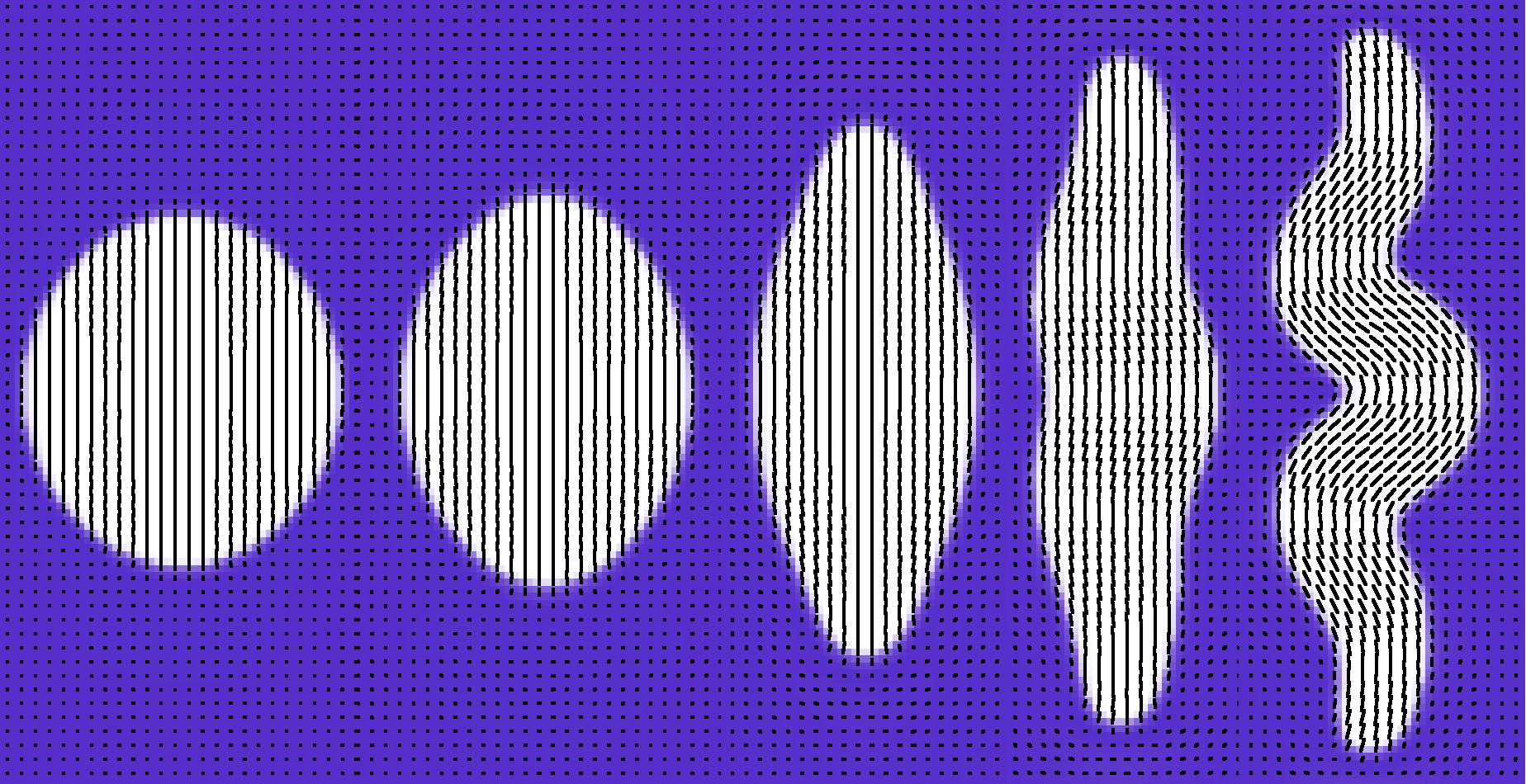}
\caption{Evolution with time of an initially circular drop of extensile active fluid. The nematic phase is white, and the isotropic phase is blue. The black lines show the scaled director $S\mathbf{n}$. }
\label{fig:drop}
\end{figure}

As a first illustration of the consequences of active stress in a lyotropic nematic, we show, in Fig.~\ref{fig:drop}, how a circular, extensile, active drop evolves with time. The drop extends parallel to the director field, and the surface alignment becomes predominantly parallel to the interface.

We emphasise that, while many models incorporate anchoring (a preferred orientation of the director field) at the interface via coupling terms in the free energy~\cite{Sulaiman2006,Patricio2011,Kopf2013}, such terms are not included in Eqn.~(\ref{eqn:freeEnergy}). Thus, our system does not exhibit any anchoring of thermodynamic origin (in contrast to the active drops in \cite{GiomiDeSimone}). Rather, the active stresses alone generate a preferential orientation - a phenomenon that we shall therefore term `active anchoring'. From Eq.~(\ref{eqn:navierStokes}), we identify the active force density as $\nabla \cdot (-\chi \mathbf{Q})$. Denoting the unit normal to the interface (pointing out of the nematic region) as $\mathbf{m}$, the force density is
\begin{eqnarray}
\mathbf{F}^{\text{active}}&=&\chi \{\vert\nabla S\vert\left(2(\mathbf{m.n})\mathbf{n}-\mathbf{m}\right)\nonumber \\
&-& 2S\left(\mathbf{n}(\mathbf{\nabla.n})+(\mathbf{n.\nabla})\mathbf{n}\right) \}\;.  \label{eqn:interfaceForce}
\end{eqnarray}
This active force has contributions from gradients in both the nematic order and orientation. The latter may be neglected if distortions in the director are small. In this approximation, the components of $\mathbf{F}^{\text{active}}$ perpendicular and parallel to the interface area are
\begin{align}
F_{\perp}^{\text{active}}&=\chi\vert\nabla S\vert\left(2(\mathbf{m.n})^{2}-1\right), \label{eqn:interfaceForcePerp}   \\
F_{\parallel}^{\text{active}}&=2\chi\vert\nabla S\vert(\mathbf{m.n})(\mathbf{l.n}),  \label{eqn:interfaceForcePara}
\end{align}
where $\mathbf{l}$ is the unit vector tangent to the interface. Both components contribute to active anchoring, but in different ways. First we consider the normal force component. From Eq.~(\ref{eqn:interfaceForcePerp}), we find that $F_{\perp}^{\text{active}}=\chi\vert\nabla S\vert$ where $\mathbf{m}$ and $\mathbf{n}$ are parallel, and $F_{\perp}^{\text{active}}=-\chi\vert\nabla S\vert$ where they are perpendicular. Thus, for extensile activity, the drop is extended where the interfacial alignment is homeotropic, and compressed where it is planar, causing an initially circular drop to be stretched along the nematic director as shown in Fig.~\ref{fig:drop}. As a result, the director field is oriented parallel to the interface everywhere except at the ends of the elongated structure. In the case of a contractile suspension, the forces are reversed, so that the nematic drop extends perpendicular to the nematic director, corresponding to homeotropic active anchoring.

\begin{figure}
\centering
\includegraphics[width=\linewidth]{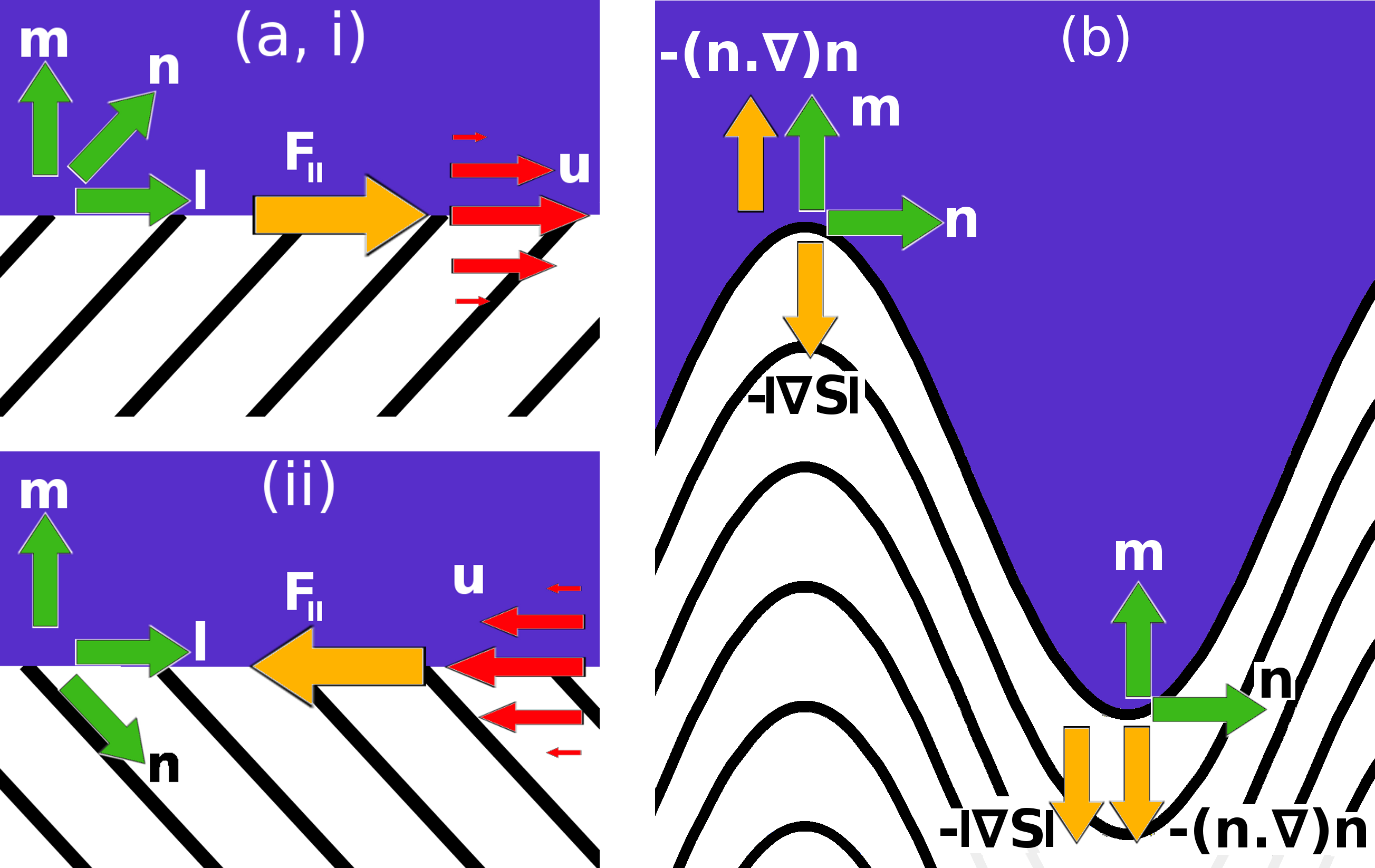}
\caption{(a) Schematic illustration showing the tangential force and resulting flow field for different director orientations (thick black lines) at the interface. (b) Schematic illustration of the forces acting at a curved interface.}
\label{fig:schematics}
\end{figure}

However, elongation is not the sole cause of active anchoring, as can be seen by considering the tangential force,  Eq.~(\ref{eqn:interfaceForcePara}). When the director is oblique to the interface (i.e. neither homeotropic nor planar), $F_{\parallel}^{\text{active}}$ will be nonzero, generating a flow along the interface. The resulting velocity gradient between the interface and the bulk nematic has a tendency to rotate the director (assuming the system is flow-tumbling), as dictated by the convective terms of Eq.~(\ref{eqn:berisEdwards}). In the extensile case, the director is in stable equilibrium when planar to the interface, as illustrated schematically in Fig.~\ref{fig:schematics}(a). In the contractile case, the flow direction is reversed, rotating the director towards the homeotropic configuration. Thus, this rotation effect acts in accord with the elongation effect to produce the active anchoring.

Planar anchoring can be observed in studies of growing bacterial colonies~\cite{Halletschek2007,Volfson2008}. We conjecture that in such systems, the division of bacteria along their long axis provides an extensile stress and hence active anchoring may provide an explanation for this behaviour.

\begin{figure}
\includegraphics[width=\linewidth]{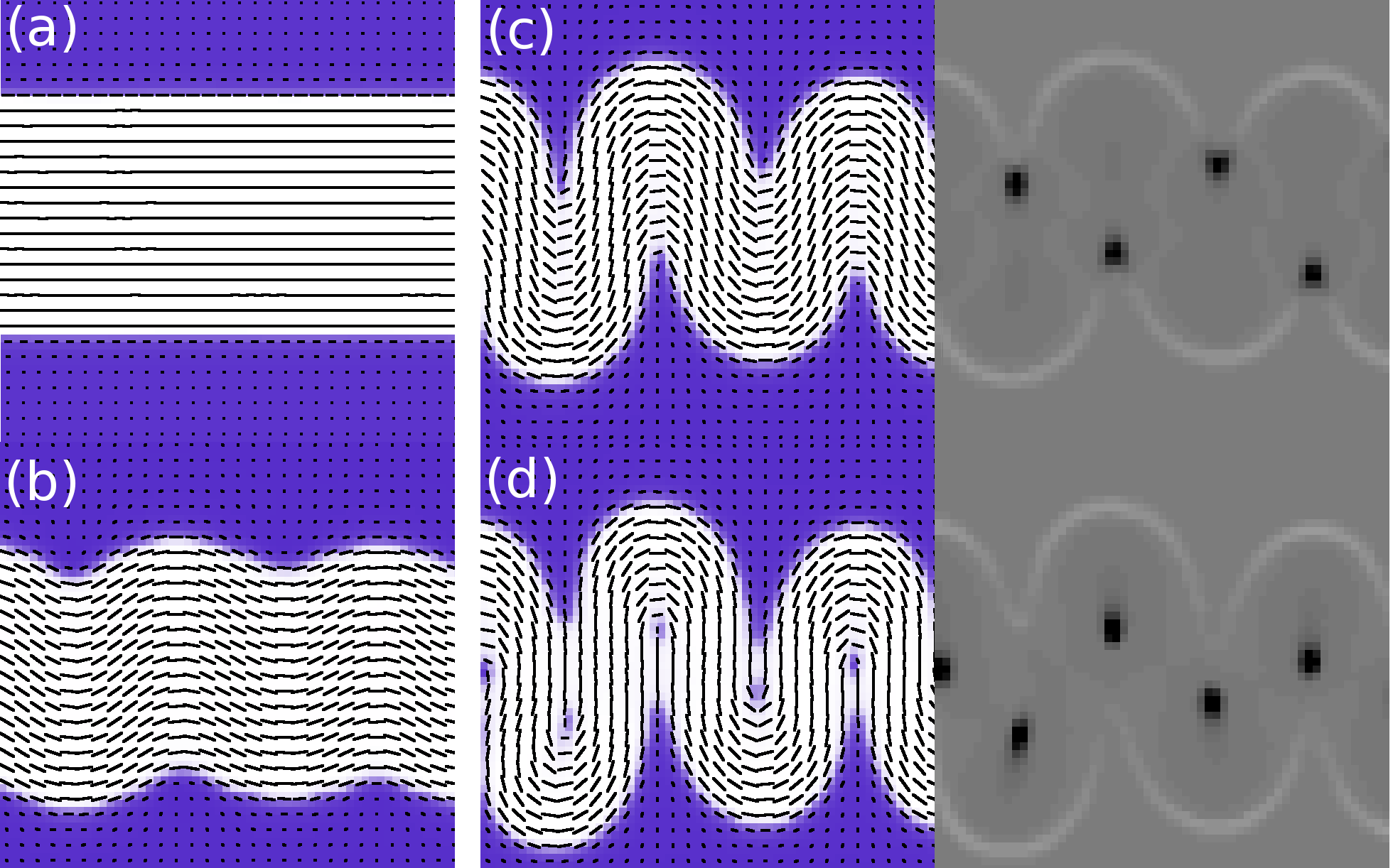}
\caption{Interface instability and defect formation in a stripe of extensile active nematic, with time running from (a) to (d). For panels (c) and (d), the rhs shows the charge density $q$, with dark (light) shades corresponding to positive (negative) charge. We omit charge density plots for (a) and (b) since there is little separation of charge at these stages.}
\label{fig:band}
\end{figure}

 In the last frame of Fig.~\ref{fig:drop}, the well-known hydrodynamic instability of an extensile active nematic to a bend deformation~\cite{Sriram2010} is starting to develop. In a bulk system this leads to active turbulence. To demonstrate its role in the behaviour of a lyotropic active fluid, Fig.~\ref{fig:band} shows the time evolution of a stripe of  extensile, active nematic. The bend instability leads to pronounced undulations of the stripe. Note the asymmetry between the rounded convex (with respect to the nematic) sections of interface and the cusp-like concave sections.

In Fig.~\ref{fig:band}(d), topological defects of charge $+1/2$ pinch off from the cusps and migrate into the nematic. This is in contrast to bulk active nematics, where topological defects of charge $+1/2$ and $-1/2$ are always produced in pairs~\cite{Dogic2012,Giomi2013,ThampiYeomans}. Here, we will need to characterise topological charge in terms of a diffuse charge density, instead of point charges. We thus define the topological charge contained within a boundary loop $\partial\mathcal{R}$.
\begin{equation}
m = \oint_{\partial\mathcal{R}} \frac{1}{8\pi}\left(Q_{x\alpha}\partial_{\beta}Q_{y\alpha}-Q_{y\alpha}\partial_{\beta}Q_{x\alpha}\right)dr_{\beta}
\end{equation}
which returns the standard definition~\cite{DeGennesBook} for $S=1$. Green's theorem gives the corresponding charge density as
\begin{equation}
q = \tfrac{1}{4\pi}\left(\partial_{x}Q_{x\alpha}\partial_{y}Q_{y\alpha} - \partial_{x}Q_{y\alpha}\partial_{y}Q_{x\alpha}\right)  \label{eqn:charge}.
\end{equation}
 Note that Eq.~(\ref{eqn:charge}) predicts that curved interfaces in general have a charge density. Assuming the director takes a constant orientation with respect to the interface, the charge density is positive for concave and negative for convex interfaces. However, it is apparent from Fig.~\ref{fig:band}(c-d,rhs) that the charge is not symmetrically distributed: the negative charge becomes widely spread over long regions of gentle curvature, while the positive charge is far more concentrated at sharp cusps. Eventually the positive point defects are pinched off, leaving behind a net negative charge at the interface.
 
\begin{figure}
\includegraphics[width=\linewidth]{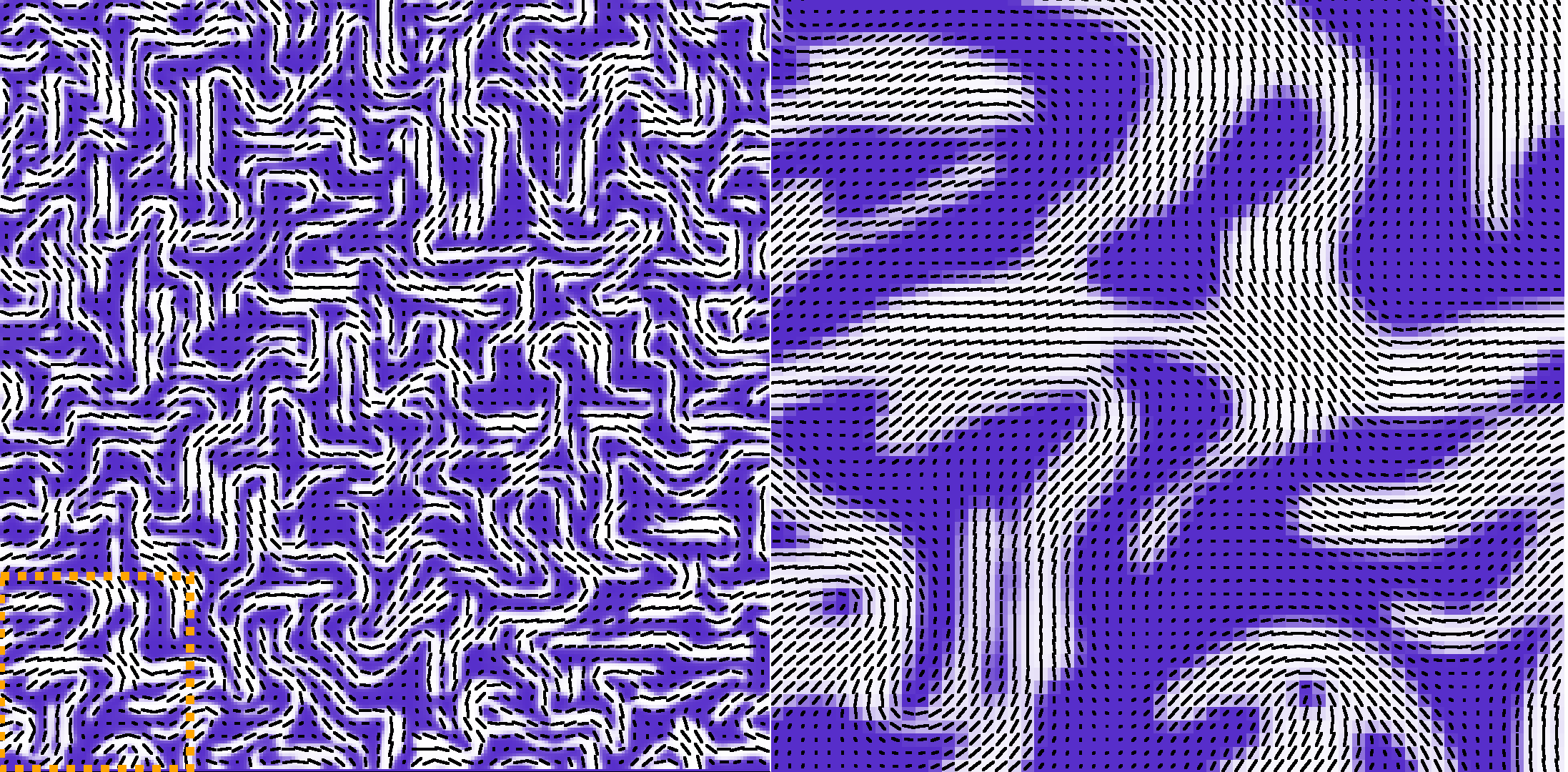}
\caption{A typical configuration of the concentration $\phi$ and the director in the steady state of a system with 50\% nematic and activity $\chi=0.005$. The rh panel is an enlargement of the orange square.}
\label{fig:decomp}
\end{figure}
To understand this asymmetry, we look again at Eq.~(\ref{eqn:interfaceForce}). Assuming active anchoring  (i.e. $\mathbf{n}$ remains perpendicular to $\mathbf{m}$ in the extensile case) and taking into account gradients of $\mathbf{n}$, the force is 
\begin{equation}
F_{\perp}^{\text{active}}=\chi\left\{-\vert\nabla S\vert\ - 2S\mathbf{m.}(\mathbf{n.\nabla})\mathbf{n} \right\}\;.  \label{eqn:interfaceForceWave}
\end{equation}
The first term represents the force arising from the gradient in nematic order, which always acts inwards. The second term may be of either sign depending on the position along the interface. As Fig.~\ref{fig:schematics}(b) illustrates, on the concave parts of the interface   $-(\mathbf{n.\nabla})\mathbf{n}$ is opposed to $\mathbf{m}$ and therefore this force contribution is directed inwards, while at the convex parts, $-(\mathbf{n.\nabla})\mathbf{n}$ is aligned with $\mathbf{m}$ and hence the force contribution is directed outwards. Thus, in the concave parts, the two force contributions combine to give a strong force that pulls the interface sharply inwards, while in the convex parts, the two contributions are opposed so the resultant force is weak. For the case of a contractile active nematic, the evolution of topological charge is the same because the change of sign of $\chi$ is cancelled out by the change in active interface alignment from planar to homeotropic.

\begin{figure}
\includegraphics[width=\linewidth]{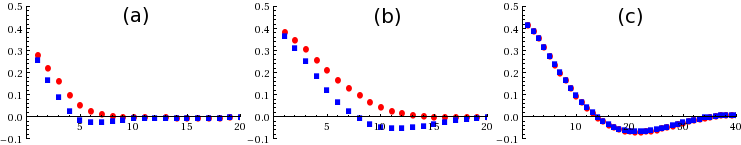}
\caption{Plots of the radial averages of the correlation functions $c_{\parallel}$ (red circles) and $c_{\perp}$ (blue squares) against radial distance $d$ for $50\%$ nematic mixtures with (a) $\chi=0.005$ (extensile) at the steady state, (b) $\chi=0.0001$ at the steady state, and (c) $\chi=0$ (passive) at a point in time during the phase separation process.}
\label{fig:correlations}
\end{figure}

When a uniformly mixed state of a lyotropic nematic is quenched to below its ordering temperature, the free energy in Eq.~(\ref{eqn:freeenergy}) drives it to phase-separate into nematic and isotropic regions. However, activity disrupts and limits the phase ordering. Nematic regions are carried around by the flow, dynamically colliding, breaking up, coalescing and being re-formed, and a steady state is reached where the domains size saturates. This is illustrated in Fig.~\ref{fig:decomp} where we show a typical configuration for a system of size $480\times 480$ (only a $240\times 240$ portion is shown) lattice spacings with periodic boundary conditions, activity $ \chi=0.005 $ and equal concentrations of the isotropic and nematic phases. A movie showing the evolution of the fields is available in the S.I.~\cite{SIvideo}. 

To quantify the characteristic domain size of the system, we construct correlation functions. The elongation effect suggests that we need two such functions, separately measuring the correlations parallel and perpendicular to the director. To this end, we define,
\begin{align}
c_{\parallel}(\mathbf{d}) &= \frac{1}{N}\sum\limits_{\mathbf{r}}\tfrac{1}{2}g(\mathbf{r},\mathbf{r}+\mathbf{d})\left(\hat{d}_{\alpha}\hat{d}_{\beta}Q_{\alpha\beta}(\mathbf{r})+1\right),\\
c_{\perp}(\mathbf{d}) &= \frac{1}{N}\sum\limits_{\mathbf{r}}\tfrac{1}{2}g(\mathbf{r},\mathbf{r}+\mathbf{d})\left(\hat{d}_{\alpha}\hat{d}_{\beta}\epsilon_{\alpha\kappa}\epsilon_{\beta\lambda}Q_{\kappa\lambda}(\mathbf{r})+1\right)
\end{align}
where $\mathbf{r}$ are the lattice nodes of the simulation, $N$ is the total number of nodes, $\hat{\mathbf{d}}$ is the unit vector along the displacement vector $\mathbf{d}$, and
\begin{equation}
g(\mathbf{r},\mathbf{r}')=\left(2\phi(\mathbf{r})-1\right)\left(2\phi(\mathbf{r}')-1\right).
\end{equation}

Radial averages of $c_{\parallel}$ and $c_{\perp}$ are plotted for an extensile system with activity $\chi=0.005$ in Fig.~\ref{fig:correlations}(a). $c_{\parallel}$ decays more slowly than $c_{\perp}$, which shows anticorrelations at medium distances. This is in agreement with the elongated domains that we see in Fig.~\ref{fig:decomp}.  Fig.~\ref{fig:correlations}(b) shows an extensile system with weaker activity. The correlation lengths are longer, indicating that the lower stirring of the system allows for the formation of larger domains, but the anisotropy remains. Fig.~\ref{fig:correlations}(c) compares a passive system at  a moment during the phase ordering process confirming that there is no domain anisotropy relative to the orientation of the director.

\begin{figure}
\includegraphics[width=\linewidth]{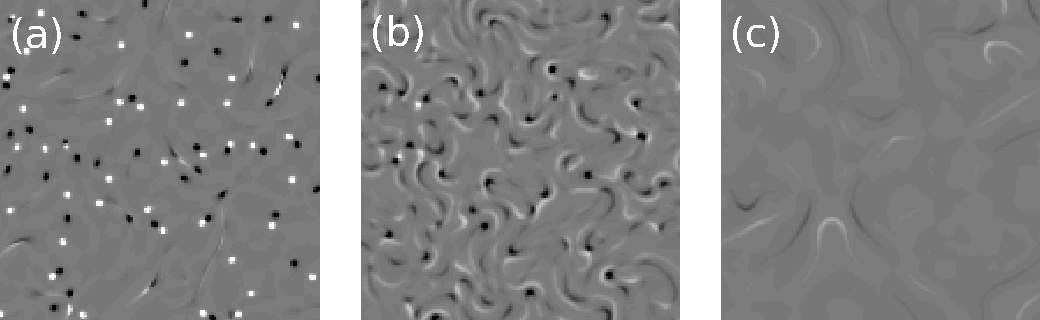}
\caption{Charge density $q$ shown for a $275\times 275$ area of the simulation box for (a) 100\% nematic with $\chi=0.005$, (b) 50\% nematic with $\chi=0.005$, (c) 50\% nematic with $\chi=0$.}
\label{fig:chargeDistribution}
\end{figure}

Another way in which the active lyotropic mixture is distinguished from both the passive lyotropic mixture and a pure active nematic is in the distribution of topological charge. In the case of a pure active nematic, with no isotropic regions, topological charge evolves through the creation and annihilation of pairs of $+1/2$ and $-1/2$ defects~\cite{Giomi2013,ThampiYeomans}. This is shown in Fig.~\ref{fig:chargeDistribution}(a), where charge is concentrated, with positive (dark) and negative (light) point charges equally abundant. Fig.~\ref{fig:chargeDistribution}(b) shows a system with the same value of $\chi$, but with 50\% nematic. A large number of positive point charges remain, but the negative charge is predominantly smeared along the interfaces. For comparison, Fig.~\ref{fig:chargeDistribution}(c) shows that there is little separation of charge in the case of a passive fluid undergoing phase separation.

To summarise, we have shown that active stresses at the interface between an active nematic and an isotropic fluid lead to domain elongation, effective anchoring and the asymmetric production of topological defects. Moreover, active stirring causes phase-separating mixtures to reach a steady state characterised by finite domain lengths. To assess the robustness of these results, in the S.I. we present simulations for different system parameters, and find that the phenomena observed in Figs.~\ref{fig:drop} and \ref{fig:band} are qualitatively unchanged. We hope that our predictions will help to motivate and explain experiments on systems as diverse as bacterial colonies, crowded microswimmers and driven microtubule suspensions.

We thank Guillaume Duclos, Daniel J. Needleman, David E. Nelson, Wilson C. K. Poon, Pascal Silberzan, Nuno M. Silvestre and Margarida M. Telo da Gama for enlightening discussions. M.L.B. acknowledges funding from the Portuguese Foundation for Science and Technology (FCT) through grants SFRH/BPD/73028/2010, EXCL/FIS-NAN/0083/2012 and PEst-OE/FIS/UI0618/2014. S.P.T. and J.M.Y. acknowledge funding from the ERC Advanced Grant MiCE.

\bibliography{refe.bib}

\renewcommand\thefigure{S\arabic{figure}} 
\setcounter{figure}{0}

\section*{Supplementary information}

In the main paper we use a single set of simulation parameters, namely $S_{\mathrm{nem}}=1$, $A=0.08$, $C=0.5$, $L=0.005$, $K=0.01$, $\Gamma=0.1$, $M=0.1$, $\eta=1/6$, $b=0.1$, $\zeta=0.3$, $\rho=40$ and $\chi=0.005$. Here we consider the effect of changing some of these parameters. The first variation we consider is a five-fold increase in $A$ and $K$ to $A=0.4$ and $K=0.05$. This has the effect of increasing the interfacial tension between the two fluid phases while leaving the interfacial width approximately unchanged. The second variation is a ten-fold decrease in the mobility parameters, such that $\Gamma=0.01$ and $M=0.01$. This slows down the rate of approach to equilibrium, and hence weakens the role of thermodynamics compared to flow. And third, we consider removing the substrate friction by setting $b=0$.\\

\begin{figure}
\includegraphics[width=\textwidth]{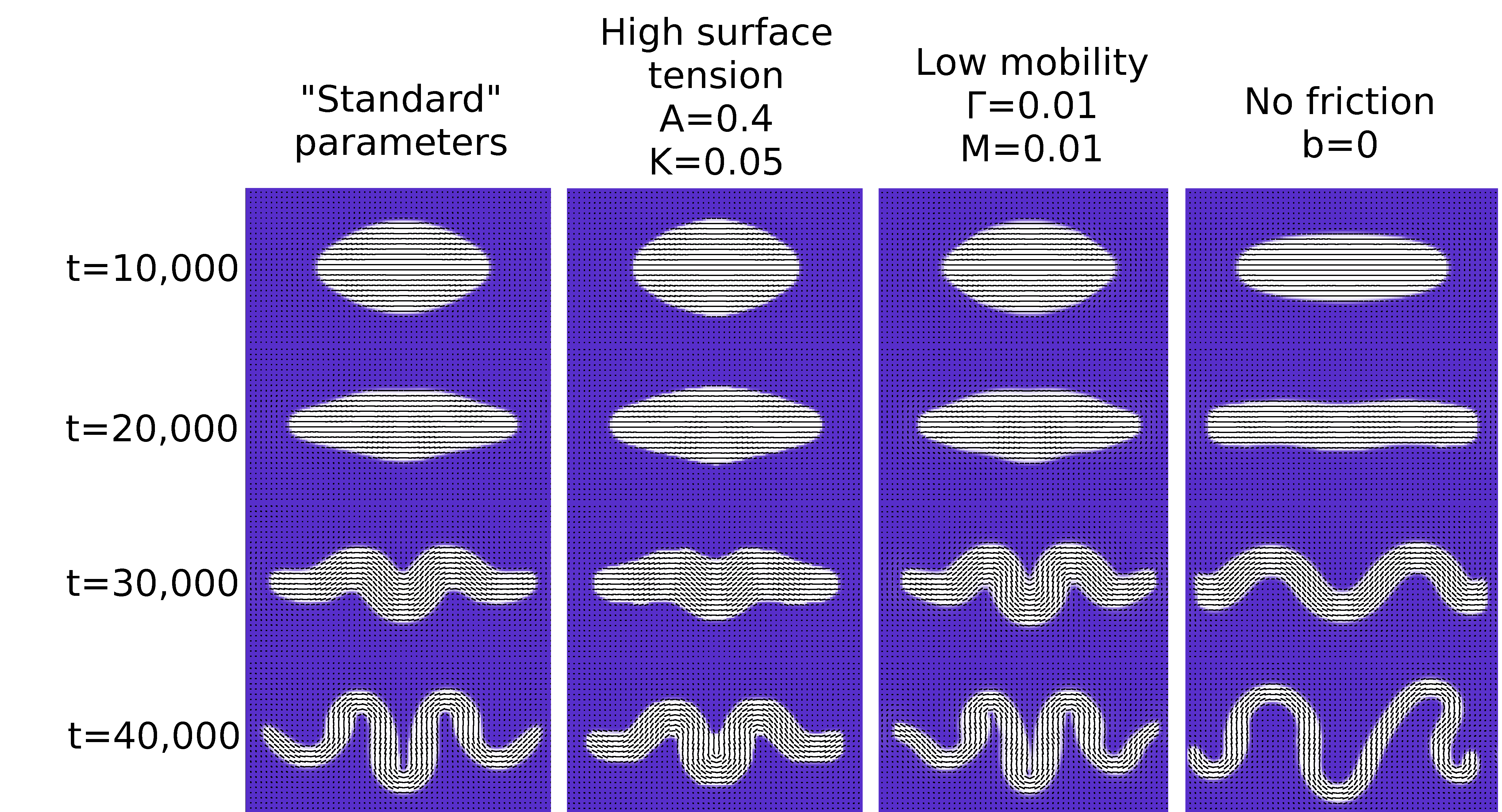}
\caption{Evolution in time of an initially circular drop of extensile active fluid for the various parameter sets.}
\label{fig:SIdrop}
\end{figure}

In figure~\ref{fig:SIdrop}, we compare, for the different parameters, the evolution of an initially circular drop, as shown in Fig.~1 of the main paper. Although there is variation in the exact drop shape and the rate at which it evolves, all four parameter sets show the same qualitative stages of behaviour: elongation, active anchoring, and the eventual undulatory instability.\\

\begin{figure}
\includegraphics[width=\textwidth]{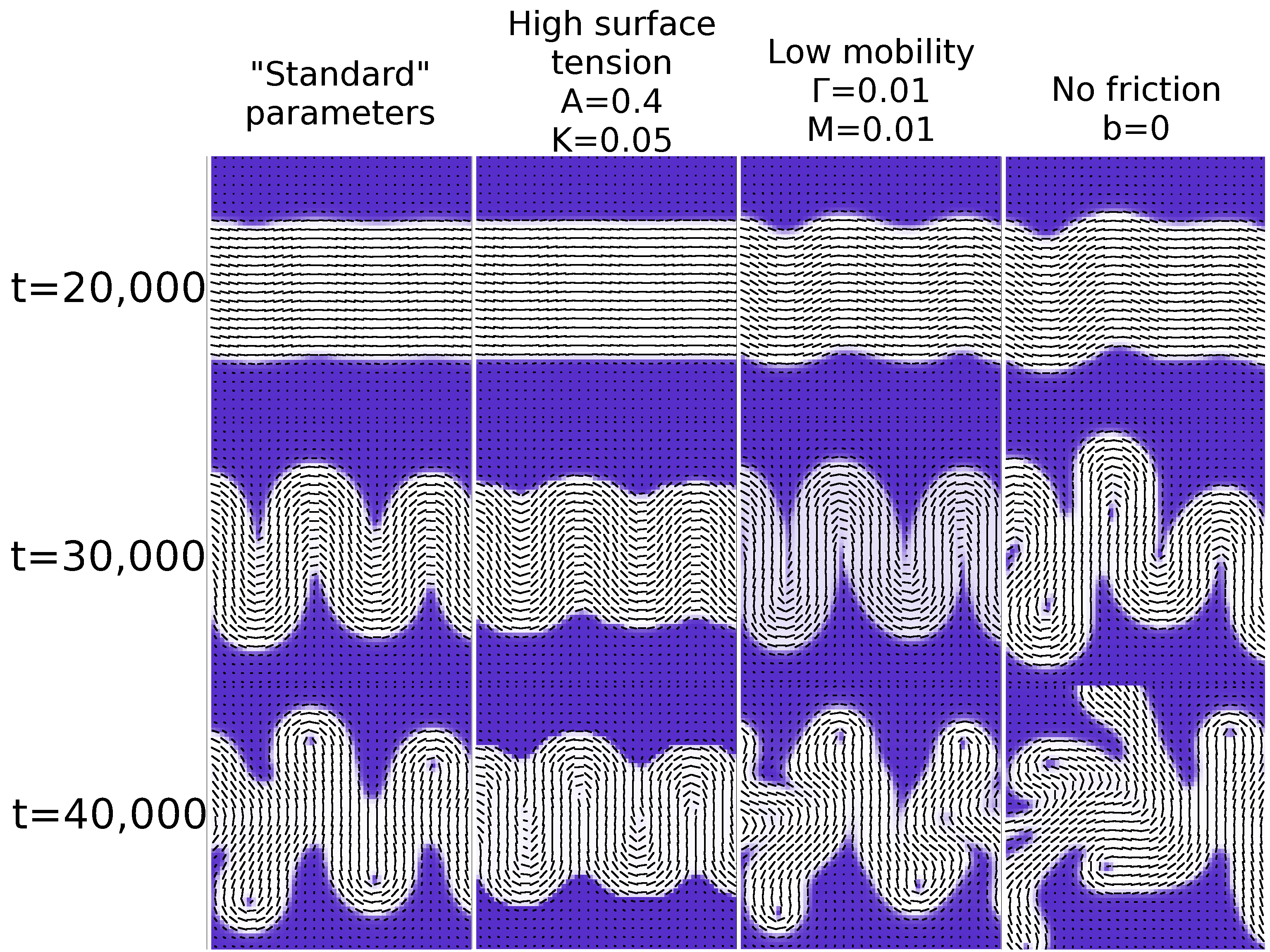}
\caption{Evolution in time of a band of extensile active fluid for the various parameter sets.}
\label{fig:SIband}
\end{figure}

In figure~\ref{fig:SIband}, we compare the evolution of an initially straight stripe, as investigated in Fig.~3 of the main paper. Again, the four situations differ in the details; e.g. in the high surface tension case the amplitude of the interface undulations is smaller and the point defects do not have an isotropic core, while the frictionless case evolves more quickly. But notwithstanding these differences, the same qualitative stages are observed. The stripe undergoes a waving instability, in which the concave parts of the interface form sharp cusps while the convex parts gain only a small curvature. Finally, singular point defects, of charge +1/2, are emitted from the interface into the bulk nematic.

\end{document}